\def\year{2021}\relax
\documentclass[letterpaper]{article} 
\usepackage{aaai21}  
\usepackage{times}  
\usepackage{helvet} 
\usepackage{courier}  
\usepackage[hyphens]{url}  
\usepackage{graphicx} 
\usepackage{amsmath}
\usepackage{natbib}  
\usepackage{caption} 
\title{How Political is the Spread of COVID-19 in the United States? \\ An Analysis using Transportation and Weather Data}

\author {
        Karan Vombatkere, \textsuperscript{\rm 1}
        Hanjia Lyu, \textsuperscript{\rm 1}
        Jiebo Luo \textsuperscript{\rm 1} \\
}
\affiliations {
    \textsuperscript{\rm 1} University of Rochester \\
    kvombatk@u.rochester.edu, hlyu5@ur.rochester.edu, jluo@cs.rochester.edu
}

\begin{document}

\maketitle

\begin{abstract}
 We investigate the difference in the spread of COVID-19 between the states won by Donald Trump (Red) and the states won by Hillary Clinton (Blue) in the 2016 presidential election, by mining transportation patterns of US residents from March 2020 to July 2020. To ensure a fair comparison, we first use a K-means clustering method to group the 50 states into five clusters according to their population, area and population density. We then characterize daily transportation patterns of the residents of different states using the mean percentage of residents traveling and the number of trips per person. For each state, we study the correlations between travel patterns and infection rate for a 2-month period before and after the official states reopening dates. We observe that during the lock-down, Red and Blue states both displayed strong positive correlations between their travel patterns and infection rates. However, after states reopened we find that Red states had higher travel-infection correlations than Blue states in all five state clusters. We find that the residents of both Red and Blue states displayed similar travel patterns during the period post the reopening of states, leading us to conclude that, on average, the residents in Red states might be mobilizing less safely than the residents in Blue states. Furthermore, we use temperature data to attempt to explain the difference in the way residents travel and practice safety measures.
\end{abstract}

\section{Introduction}
There have been many analyses done recently on investigating the spread of COVID-19 in relation to the state policies, political ideology regarding travel guidelines, mask-usage and state-specific reopening rules. Both news media~\footnote{https://www.cnn.com/2020/07/08/politics/what-matters-july-8/index.html} and academic researchers~\cite{rothgerber2020politicizing, blagov2020adaptive} have intended to investigate the relationship between the difference of infection growth and the politicization across residents' behaviours and psychological factors.\\

\begin{figure}[htbp]
\centering
\includegraphics[width=0.9\columnwidth]{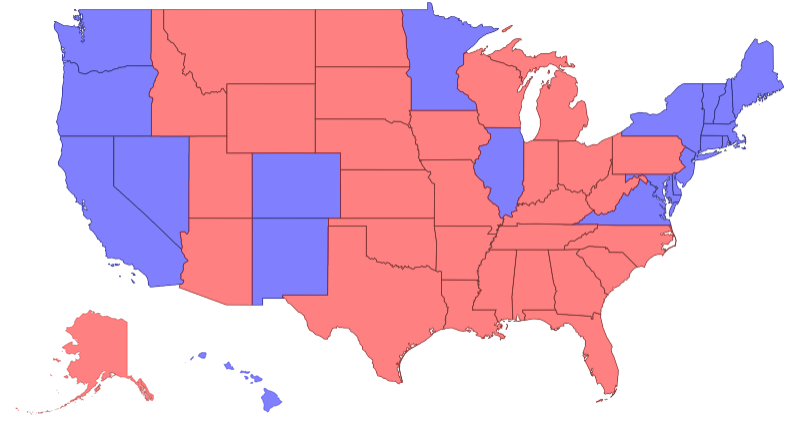}
\caption{2016 US presidential election results.}
\label{election2016map}
\end{figure}

\begin{figure}[htbp]
\centering
\vspace{-0.2cm}
\includegraphics[width=0.9\columnwidth]{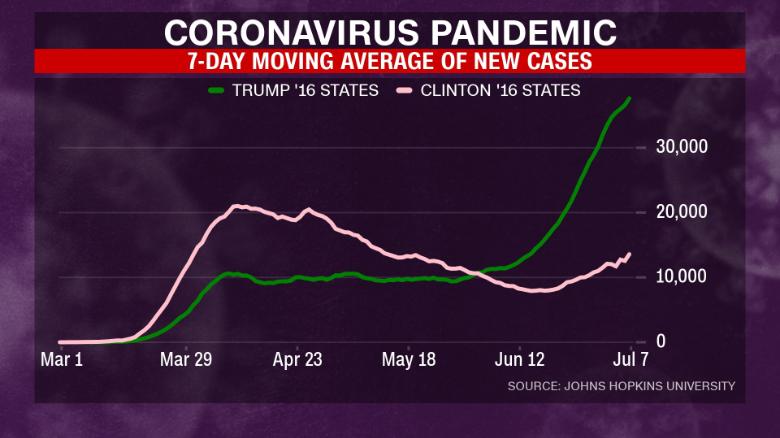}
\caption{Evolution of daily confirmed cases (Red vs. Blue States). Source: CNN.}
\label{trump_hillary_cases}
\end{figure}

Figure~\ref{election2016map} shows a map of US states with the results of the 2016 presidential election, where Hillary Clinton won 21 (Blue) states and Donald Trump won 30 (Red) states. We intend to study the spread of COVID-19 in these two sets of states by mining transportation patterns and attempting to correlate them with the daily infection rate. The graph of daily confirmed infections in Trump vs. Clinton states shown in Figure~\ref{trump_hillary_cases} was publicized by CNN\footnote{https://www.cnn.com/2020/07/08/politics/what-matters-july-8/index.html} in July 2020 and caught our attention. The goal of the analysis in this paper is to use transportation patterns along with weather data to provide insight into why COVID-19 cases might be growing at a higher rate in the Red states than the Blue states, after states started to officially reopen in May 2020.

\begin{figure}[htbp]
\centering
\includegraphics[width=0.9\columnwidth]{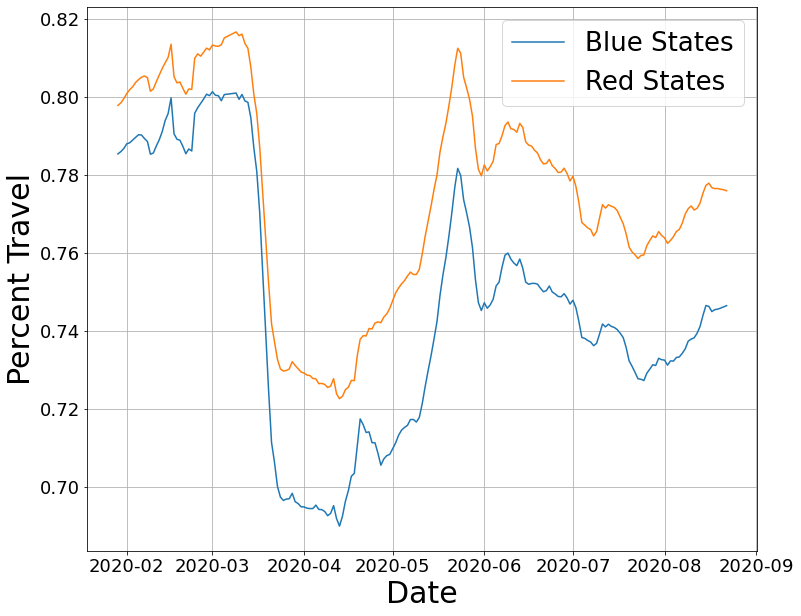}
\caption{7-Day moving average of daily percent travel (Red vs. Blue states)}
\label{percent_travel}
\end{figure}

Preliminary analysis of the transportation data indicates that both the Red and Blue states showed a sharp decrease in travel once the official stay-at-home order was passed on March 13, 2020. Figure~\ref{percent_travel} shows the 7-day moving average of the mean percentage of residents traveling daily from February 2020 to August 2020. We see a relatively consistent travel pattern that correlates with the increases and decreases in travel based on other national guidelines, such as the opening of businesses such as restaurants and bars. While a higher percentage of residents of the Red states appear to travel daily, it is important to note that this was also the case prior to COVID-19. The slight increase in the difference of travel between the Red and Blue states after the stay-at-home order was passed is marginal (about ~2\%) and was not shown to correlate with the much larger difference in infection growth rates.

There are a variety of pandemic growth models that have been studied, and almost all models involve exponential growth. Exponential growth can have vastly difference outcomes depending on the starting conditions and the exponential infection growth factor, and it is known that the infection growth depends on several factors such as travel guidelines, mask-usage, social-distancing, quarantine measures, indoor transmission~\cite{courtemanche2020strong}. Our approach to studying the spread of COVID-19 focuses primarily on transportation data: we examine the percentage of residents traveling and the number of trips taken per person, as well as the number of trips under 25 miles per person and number of trips under 10 miles per person.\\

Our contributions are summarized as follows:
\begin{itemize}
    \item We use transportation data to analyze the difference in the spread of COVID-19 between the states won by Donald Trump (Red) and the states won by Hilary Clinton (Blue) in the 2016 presidential election.
    \item We study the correlations between travel patterns and infection rates before and after the states' reopening dates.
    \item We discover that there are differences in the correlations before and after the states reopened, and temperature might have affected how people traveled.
\end{itemize}

\section{Related Work}
Data from two MTurk studies with US respondents revealed an ideological divide in adherence to social distancing guidelines during the COVID-19 pandemic~\cite{rothgerber2020politicizing}. Studies have been conducted to find the relationship between personality and the compliance of the social distancing order~\cite{blagov2020adaptive}. The last eight stay-at-home orders were issued by Republican leaders.\footnote{https://fivethirtyeight.com/features/democratic-and-gop-governors-enacted-stay-at-home-orders-on-the-same-timeline-but-all-holdouts-are-republicans/} Some elected Republican officials encouraged their constituents to patronize restaurants and bars precisely when federal health officials urged the opposite.\footnote{https://www.nytimes.com/2020/03/15/us/stay-home-go-out-coronavirus.html} Political ideology represents shared beliefs, opinions, and values held by an identifiable group or constituency~\cite{freeden2004reassessing,knight2006transformations}, and it endeavors to describe and interpret the world and envision the world as it should be~\cite{jost2009political}.\\

Many studies have been done to investigate the relationship between travel restrictions and the spread of COVID-19. \citet{chinazzi2020effect} used a global metapopulation disease transmission model to project the impact of travel limitations on the national and international spread of the pandemic. \citet{linka2020outbreak} found that an unconstrained mobility would have significantly accelerated the spreading of COVID-19.\\

In our investigation, we use publicly available transportation data provided by the US Department of Transportation to study the transportation patterns of state residents. We use the John Hopkins COVID-19 reporting data to study the infection rate for each state.

\section{Data Collection}
We use two primary data sources for the analysis conducted in this paper. We also augment our analysis with data from a couple of other sources, which are mentioned below.
\subsection{Department of Transportation Data}
This data set is publicly available and can be accessed through the Socrata API\footnote{https://data.bts.gov/resource/w96p-f2qv.json}. The travel statistics are produced from an anonymized national panel of mobile device data from multiple sources, and a weighting procedure expands the sample of millions of mobile devices, so the results are representative of the entire population in a nation, state, or county.\\

For our analysis we use the state-level transportation data, which contains rich information about the number of people travelling on a daily basis. Some of the key features in the raw data that we use are the population staying at home and the total number of trips taken of different distances. We use these to compute the following travel feature vectors, which are then normalized by population, i.e. per person basis. Trips under 25 miles and 10 miles are chosen to examine the local mobility of residents, to better capture characteristics of community transmission.

\begin{itemize}
    \item Mean percentage of the total population that traveled on a given day, $m_1(t)$
    \item Mean number of trips taken, per person, $n(t)$
    \item Number of under 10 mile trips taken, per person, $n_{10}(t)$
    \item Number of under 25 mile trips taken, per person, $n_{25}(t)$
\end{itemize}

\subsection{JHU CSSE COVID-19 Dataset}
This is the data repository for the 2019 Novel Coronavirus Visual Dashboard operated by the Johns Hopkins University Center for Systems Science and Engineering (JHU CSSE\footnote{https://github.com/CSSEGISandData/COVID-19}). We used the time series reported data to obtain the daily confirmed cases data for each state for the period from March to July 2020. We use this dataset to calculate the daily exponential infection growth rate, $r(t)$ for each state.

\subsection{Other Sources}
We leverage the US Census 2010 data\footnote{https://www.census.gov/programs-surveys/decennial-census/data.html} to obtain Area and Population data for each state. We also use data from the BallotPedia\footnote{Data available at: https://ballotpedia.org} to get the official states reopening dates used in our analysis.

\section{Methodology}
\subsection{K-Means Clustering}
We use the K-means clustering algorithm to cluster all US states based on their population, area and population density. The rationale behind using these features to cluster is to create clusters of similar states based on only the area and population, in an attempt to create a more even and fair comparison baseline for travel within the Red and Blue states. \\

The optimal number of clusters, $K$ is determined using a combination of both the Silhouette Score and the Elbow method so as to have a balance between the sum-of-squared error of the clustering and the ratio of intra-cluster to nearest-cluster distances.

\subsection{Calculating Infection Rate}
We use a simple logistic model for the infection growth rate~\cite{ma2020estimating}. The cumulative number of confirmed cases, $C(t)$ at time $t$ can be approximated by
\begin{equation}
    \frac{d}{dt}C(t)= rC(t)\large{(}1 - \frac{C(t)}{K}\large{)}
\end{equation}
where $r$ is the exponential infection growth rate of interest to us, and $K=\lim_{t\to\infty}C(t)$.\\

We see that a solution to this model for the number of confirmed cases $C(t)$ is proportional to $e^{rt}$, so in the discrete case where $t$ is measured in days we can calculate the daily infection growth rate by taking the ratio of the daily difference in cumulative confirmed cases data. For a given day $t$, the number of new cases $N(t)$ for that day can be calculated by 
\begin{equation}
    N(t) = C(t) - C(t-1)
\end{equation}

The daily exponential growth rate, $r(t)$ can then be calculated for each day by taking the ratio of successive $N(t)$, i.e. 
\begin{equation}
    r(t) = \frac{N(t)}{N(t-1)}
\end{equation}

This method is used to calculate the daily infection growth rate, $r(t)$ for each state from the cumulative confirmed COVID-19 case JHU dataset from March 2020 through July 2020. The $r(t)$ vector for each state was then used as the input to the correlation analysis conducted with the transportation pattern data.

\subsection{Correlation Analysis}
We use the Pearson correlation coefficient as a measure of the similarity between the time series data for our analysis. We set a p-value threshold of $0.05$ to ensure a $95\%$ confidence interval for all correlations reported.\\

It is well accepted that there is a leader-follower relationship between travel and infection, since the infection growth caused by travel would only manifest itself in confirmed cases data after a delay of at least 2-3 weeks based on the incubation period of COVID-19~\cite{lauer2020incubation}. Thus, in all correlation analysis we use a lag factor of 21 days to account for this incubation period and the time needed for an individual to obtain a confirmed test result. Based on empirical tests with the data with different lag factors we are able to confirm this pattern.\\

Based on the graph of daily confirmed cases in Figure~\ref{trump_hillary_cases}, we observe the daily cases for Red states start to increase in early June 2020. Since almost all states opened officially in late April - May 2020, we investigate travel patterns before and after these reopening dates, to attempt to explain the increase in spread of COVID-19 in Red states. This rationale defines our methodology for correlation analysis, and is outlined below.\\

For each state, within each state cluster, we perform the following steps. Note that a lag factor of 21 days is applied to the infection growth rate data.
\begin{enumerate}
    \item Calculate the correlations between daily travel feature vectors ($m_1(t)$, $n(t)$, $n_{10}(t)$, $n_{25}(t)$) and the infection growth rate vector $r(t)$ from March 01, 2020 to the state's official reopening date. 
    \item Calculate the correlations between daily travel feature vectors ($m_1(t)$, $n(t)$, $n_{10}(t)$, $n_{25}(t)$) and the infection growth rate vector $r(t)$ from the state's official reopening date to July 15, 2020.
\end{enumerate}

We then examine the difference between the correlation coefficients calculated by step $1$ and $2$ for Red and Blue states on a cluster-wise basis. The time period for both step $1$ and $2$ is approximately 60 days, since early May 2020 serves as the mean time during which US states officially reopened.

\section{Analysis Results}

\subsection{Clustering US States}
The K-means clustering algorithm is used to cluster the states into five clusters based on their area, population and population density. Note that Washington DC and Alaska are identified as outliers and are not included in the original clustering model, but are added back into our analysis by using the resulting model to assign them to the nearest cluster.\\

\begin{enumerate}
    \item \textbf{Cluster 1}
        \begin{itemize}
            \item \textbf{Red States:} FL
            \item \textbf{Blue States:} NY
        \end{itemize} 
     \item \textbf{Cluster 2}
        \begin{itemize}
            \item \textbf{Red States:} GA, IN, TN, MI, NC, OH, PA
            \item \textbf{Blue States:} IL, MA, NJ, VA, WA
        \end{itemize}
    \item \textbf{Cluster 3}
        \begin{itemize}
            \item \textbf{Red States:} TX
            \item \textbf{Blue States:} CA
        \end{itemize} 
      \item \textbf{Cluster 4}
        \begin{itemize}
            \item \textbf{Red States:} WV, SC
            \item \textbf{Blue States:} CT, DC, DE, HI, MD, ME, NH, RI, VT
        \end{itemize}
      \item \textbf{Cluster 5}
        \begin{itemize}
            \item \textbf{Red States:}AK, AL, AR, AZ, IA, ID, KS, KY, LA, MO, MS, MT, ND, NE, OK, SD, UT, WI, WY 
            \item \textbf{Blue States:} CO, MN, NM, NV, OR
        \end{itemize}
\end{enumerate}

\begin{figure}[h]
\centering
\includegraphics[width=1.0\columnwidth]{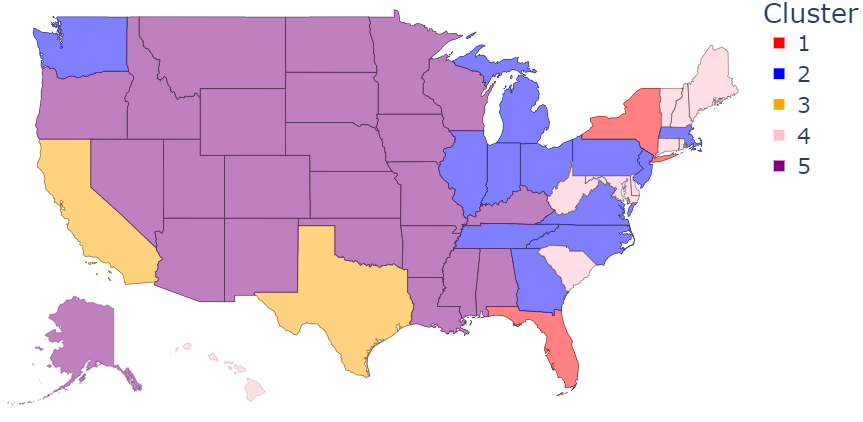}
\caption{K-Means clustering of US states into $k=5$ clusters.}
\label{state_cluster_map}
\end{figure}

The choice of $k=5$ optimal clusters is empirically determined by using the Silhouette Score and the Elbow Method together to find an appropriate grouping of states, and this clustering is visualized on a map in Figure~\ref{state_cluster_map}.

\subsection{Correlation Analysis}
Correlation analysis is done between the mined transportation patterns and the COVID-19 infection rate for each state. We examine the difference in correlations between Red and Blue states within each cluster for an approximate two month period before and after each state's reopening date. \\

\subsubsection{March 1, 2020 to States Reopening Dates}
\begin{figure}[h]
\centering
\includegraphics[width=1.0\columnwidth]{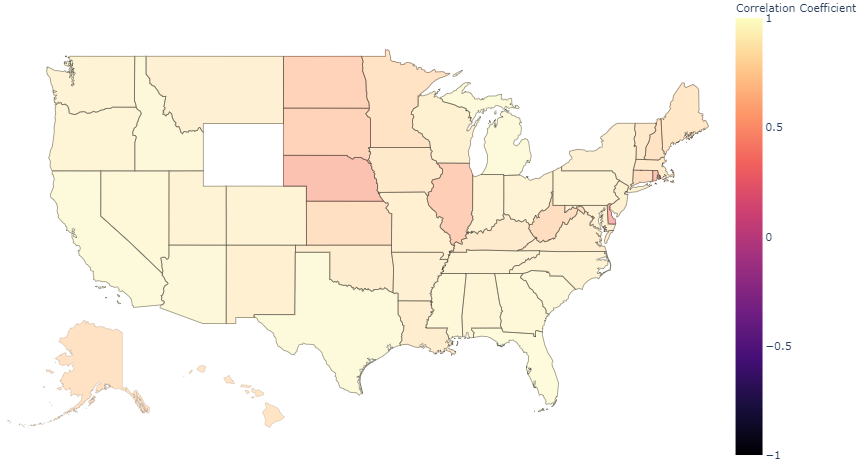}
\caption{Correlation between $m_1(t)$ and $r(t)$ from March 1, 2020 to states reopening dates.}
\label{meanpercent_preopening_map}
\end{figure}

\begin{figure}[h]
\centering
\includegraphics[width=0.85\columnwidth]{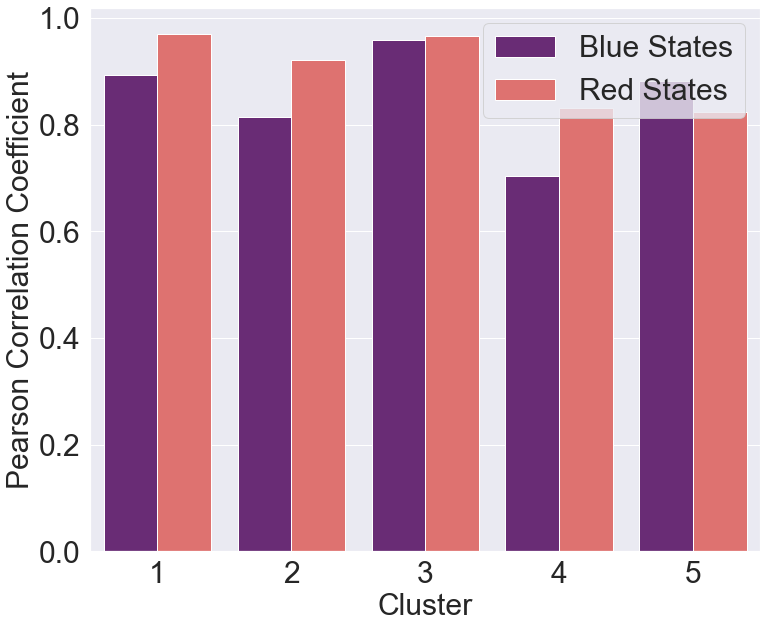}
\caption{Cluster-wise correlations between $m_1(t)$ and $r(t)$ from March 1, 2020 to states reopening dates.}
\label{meanpercent_preopening_bar}
\end{figure}

It is observed that during the quarantine period from March 1, 2020 to the states reopening dates, Red and Blue states both displayed strong positive correlations between their travel feature vectors: $m_1(t)$, $n(t)$, $n_{10}(t)$, $n_{25}(t)$ and infection growth rate vectors, $r(t)$.\footnote{The correlation coefficient of WY is the only one that is not statistically significant. ($p > 0.05$)} Figure~\ref{meanpercent_preopening_map} shows a geographic distribution of the correlations between $m_1(t)$ and $r(t)$ across all US states, and the bar plot in Figure~\ref{meanpercent_preopening_bar} visualizes this relationship for the mean value of the correlation within each of the five state clusters.\\

\begin{figure}[h]
\centering
\includegraphics[width=0.98\columnwidth]{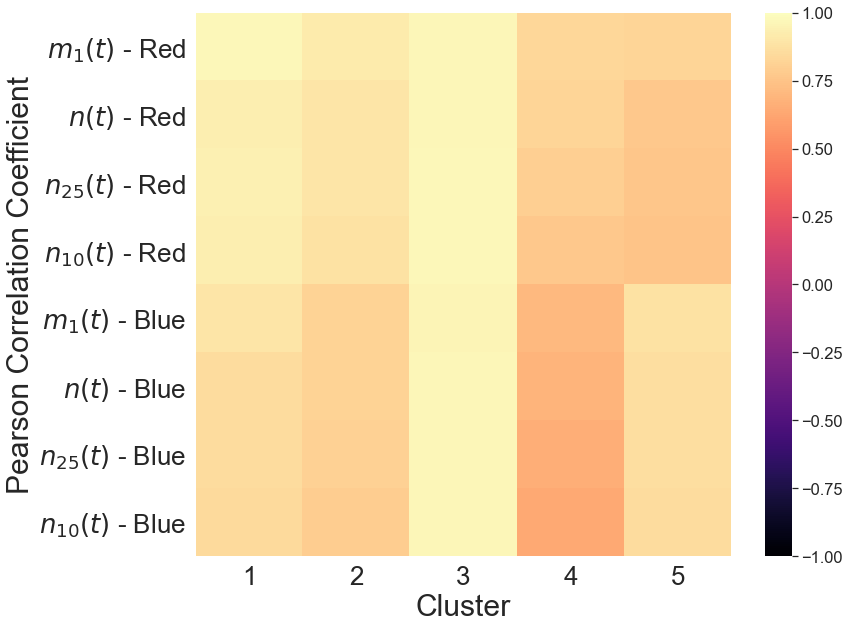}
\caption{Heat-map of correlations between travel feature vectors and $r(t)$ from March 1, 2020 to states reopening dates.}
\label{preopening_heatmap}
\end{figure}

The heat-map in Figure~\ref{preopening_heatmap} shows the consistency of the mean correlations calculated between the travel feature vectors $m_1(t)$, $n(t)$, $n_{10}(t)$, $n_{25}(t)$ and the infection growth rate $r(t)$. It becomes readily apparent that travel patterns in both Red and Blue states exhibit a strong correlation with the infection growth rate $r(t)$ across all five state clusters.\\

\subsubsection{States Reopening Dates to July 15, 2020}
\begin{figure}[h]
\centering
\includegraphics[width=0.9\columnwidth]{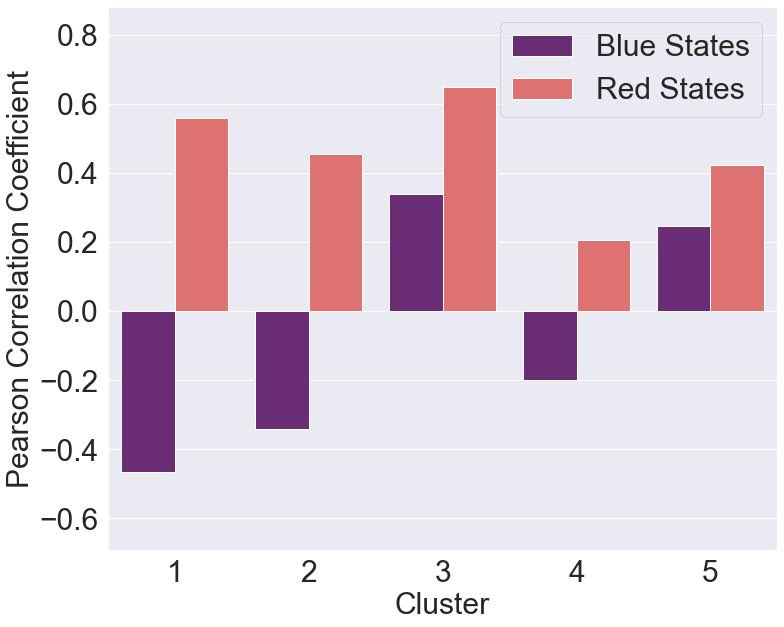}
\caption{Cluster-wise correlations between $m_1(t)$ and $r(t)$ from states reopening dates to July 15, 2020}
\label{meanpercent_postopening_bar}
\end{figure}

After states were officially reopened, correlation analysis between the travel feature vectors $m_1(t)$, $n(t)$, $n_{10}(t)$, $n_{25}(t)$ and the infection growth rate $r(t)$ yielded a very different result for Red and Blue States. These correlations are computed for the time period between the states' official reopening dates to July 15, 2020. We observe that, while travel patterns for Red States continued to display weaker positive correlations with the infection growth rate $r(t)$, there is either no correlation or negative correlations for Blue States within the same clusters. The bar plot in Figure~\ref{meanpercent_postopening_bar} captures this relationship for the mean value of the correlation within each of the five state clusters. A similar relationship is observed for the correlations of $n(t)$, $n_{10}(t)$, $n_{25}(t)$ and $r(t)$ in this time period.\\

\begin{figure}[h]
\centering
\includegraphics[width=0.98\columnwidth]{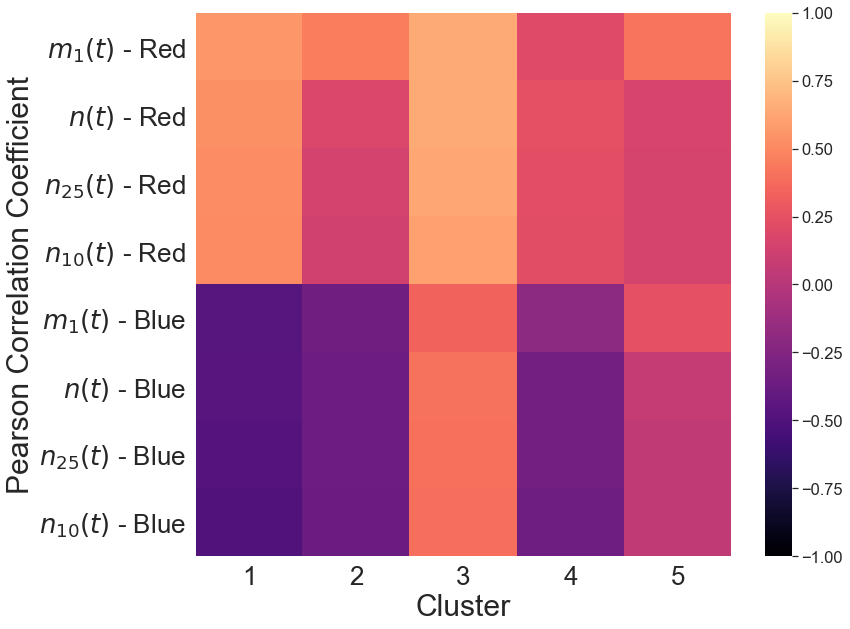}
\caption{Heat-map of correlations between travel feature vectors and $r(t)$ from states reopening dates to July 15, 2020}
\label{postopening_heatmap}
\end{figure}

The heat-map in Figure~\ref{postopening_heatmap} shows the mean correlations calculated between the travel feature vectors $m_1(t)$, $n(t)$, $n_{10}(t)$, $n_{25}(t)$ and the infection growth rate $r(t)$ for the time period between the states' official reopening dates to July 15, 2020. From the heat-map, we observe that there is a distinct difference between the correlations coefficients of the Red states and the correlation coefficients of the Blue states, for all four travel feature vectors and across all five clusters. We do see that for cluster 5, this difference is less significant. This could perhaps be attributed to the slower spread of infection in the Midwestern states due to low population density and hence lower community transmission rates.\\

\subsection{Further Analysis for Travel Patterns after States' Reopening Dates}
To better understand the difference of the correlation coefficients during the periods before and after the states reopening dates, we separate the states into four groups according to the growth pattern of the number of newly confirmed cases and the travel pattern. \\

\begin{figure}[h]
\centering
\includegraphics[width=0.95\columnwidth]{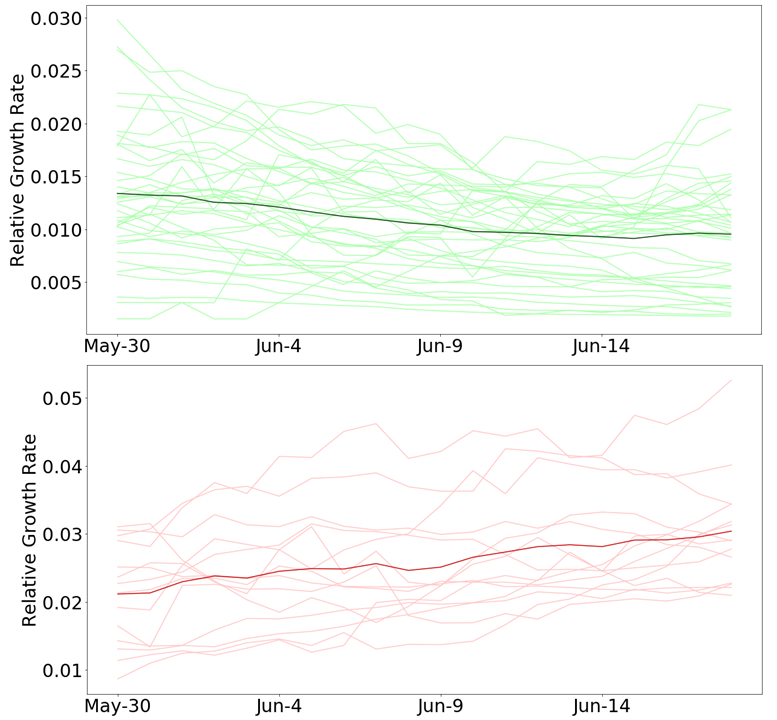}
\caption{The patterns of the relative growth rate of newly confirmed cases for the two clusters formed according to the newly confirmed cases, i.e. a cluster that was trending up and another cluster that was trending down. Note that here we do not differentiate the Red and Blue States. }
\label{pattern_relative_growth}
\end{figure}

\begin{figure}[h]
\centering
\includegraphics[width=1.0\columnwidth]{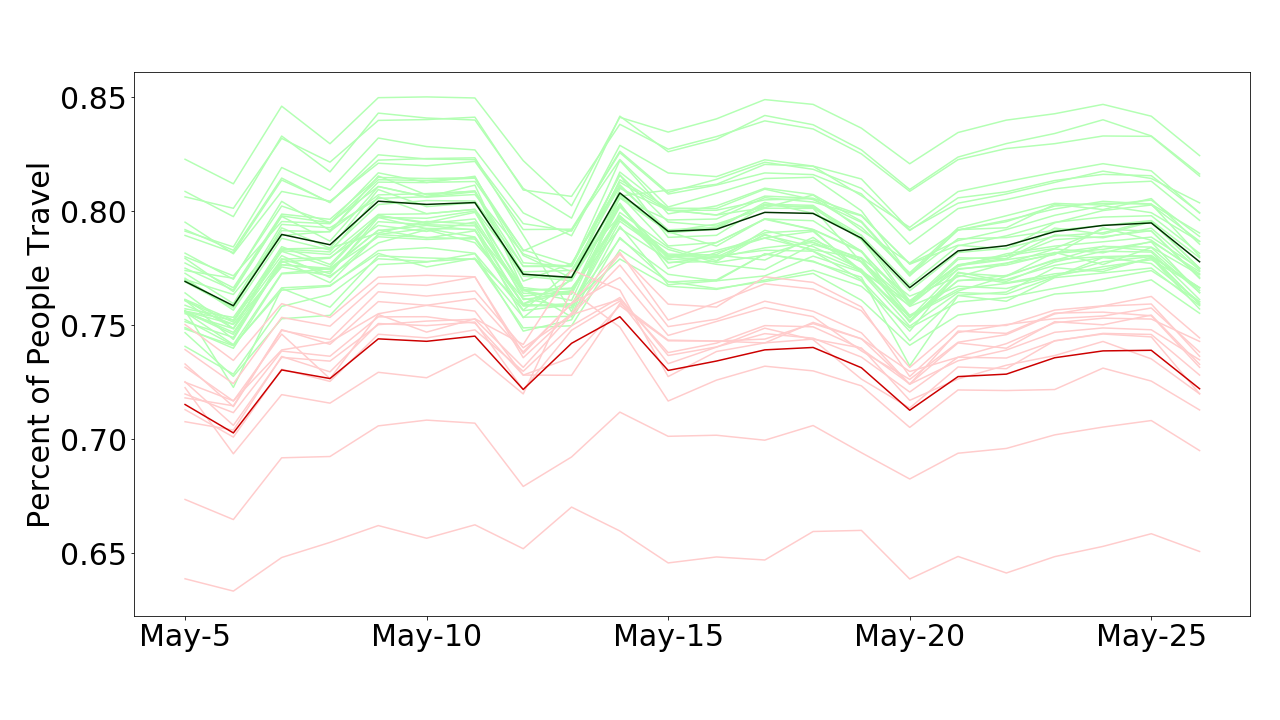}
\caption{The daily percent of people who traveled post reopening for the two clusters formed according to traveling people, i.e. a cluster with more traveling people and another cluster with fewer traveling people.  Note that here we do not differentiate the Red and Blue States. }
\label{travel_pattern}
\end{figure}

\begin{figure*}[htbp!]
\centering
\includegraphics[width=1.0\textwidth]{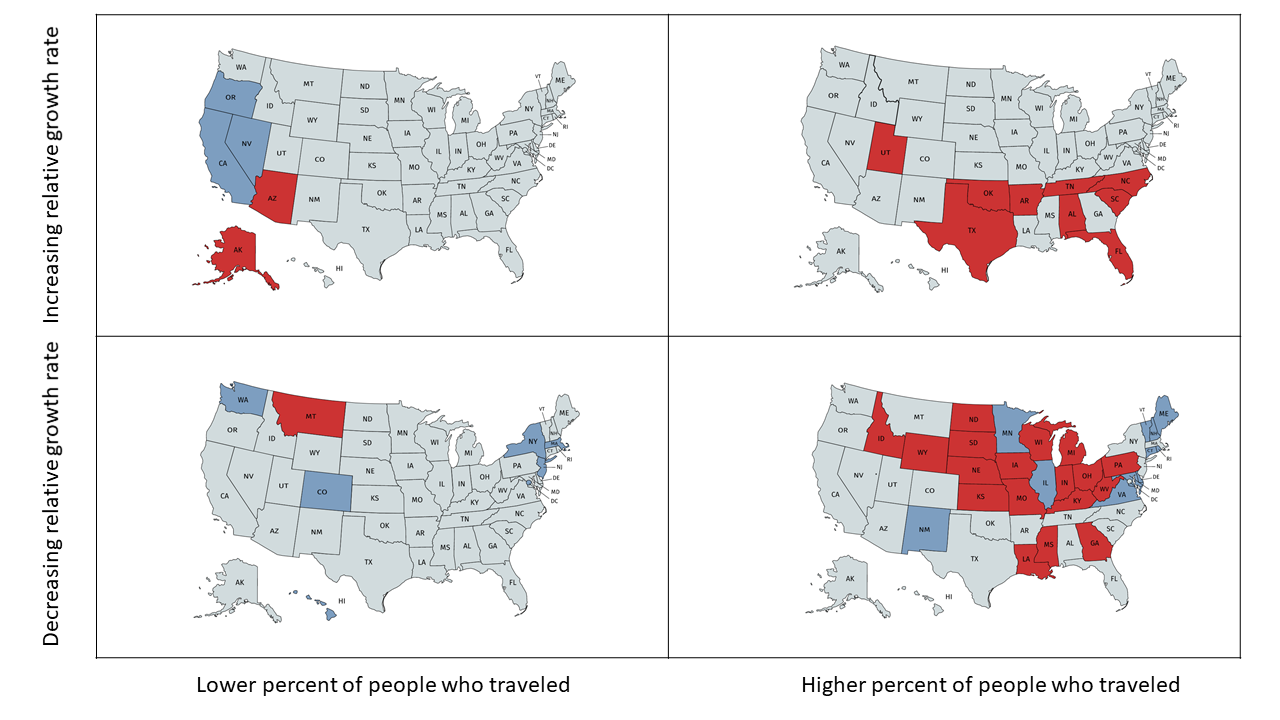}
\caption{Combination of the pattern of relative growth rate and the travel pattern.}
\label{four_groups}
\end{figure*}

We apply the K-means algorithm to the relative growth rates of the newly confirmed cases of all the states, and the highest Silhouette Score (0.522) is achieved when the number of clusters is set to two. Figure~\ref{pattern_relative_growth} shows the patterns of the relative growth rates of newly confirmed cases for both clusters. Green lines represent the states of cluster 1 and red lines represent the states of cluster 2. The dark lines represent the mean relative growth rates of each cluster. The difference of the relative growth patterns of the two clusters is clear where the relative growth rates of the newly confirmed cases of cluster 1 was trending down between May 30 and June 20, while that of the cluster 2 was trending up. Due to the difference of the pattern of relative growth rates, we separate the states into two groups: one contains the states that had a decreasing trend and the other displayed an increasing trend.\\

By applying K-means to the percent of people who traveled during reopen, we separate the states into two groups. Two clusters yield the highest Silhouette Score (0.569). Figure~\ref{travel_pattern} shows the patterns of the percent of people who traveled during reopening (roughly from May 5 to May 25). Two sets of lines represent the states that are clustered by 2-means. The dark lines are the mean percentage of each cluster. The lines for both clusters are periodic which is expected since people might travel differently on weekdays and weekends. However, on average, there are more people who traveled in the states that are represented by the green lines and fewer people in the ones that are represented by the red lines. Thus, we separate all the states into two groups based on the difference of the travel pattern.\\

Using this clustering, we split the states into four groups according to the combination of the pattern of relative growth rate and the travel pattern. Figure~\ref{four_groups} shows the splitting results. Take the bottom left section as an example: this section represents the states that had a decreasing relative growth rate and a lower percent of people who traveled post reopening. The red states are the states won by Trump in the 2016 presidential election and the blue states are the states won by Clinton. Except for the top right section where the states' relative growth rate were trending up and had a higher percent of people who traveled, the other three sections contain both Blue States and Red States. There are Blue and Red States in the top left and the bottom right sections providing an explanation for the difference of the correlation coefficients during the periods before and after the start of reopening. One the one hand, there are Red States that had a higher percent of people who traveled during reopen and meanwhile they had a decreasing relative growth rate. On the other hand, Blue States such as California, Oregon and Nevada had a lower percent of people who traveled, but they still suffered an increasing relative growth rate.\\

\begin{figure}[htbp]
\centering
\includegraphics[width=0.9\columnwidth]{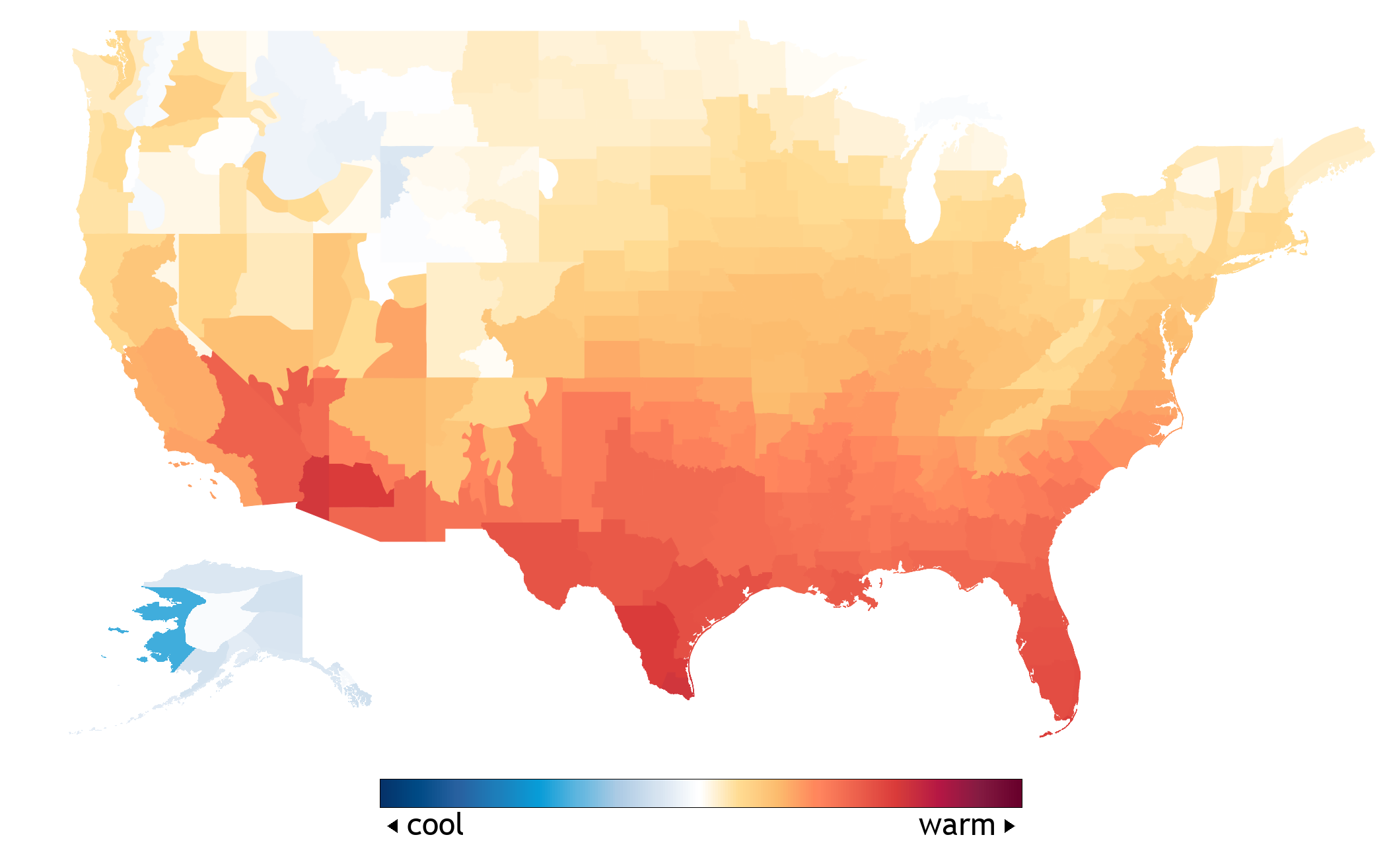} 
\caption{Average temperature by state in May, 2020.}
\vspace{-0.1cm}
\label{tem}
\end{figure}

To obtain a better insight into the patterns of the relative growth rate, we compare the states of the first row and the second row and find that most states in the first row are in the southern part of the country and the majority states in the second row are in the northern part of the country. This leads to a hypothesis that the difference of the relative growth rate could be related to the temperature. Figure~\ref{tem} shows the average temperature in May 2020\footnote{https://www.climate.gov/maps-data/data-snapshots/averagetemp-monthly-cmb-2020-05-00?theme=Temperature}. The temperature map roughly corresponds to the patterns of the relative growth rate where in the south, states were warmer and they displayed an increasing relative growth rate, while in the north, states were cooler and they had a decreasing relative growth rate.\\

Previous study has also shown that there is no evidence supporting that case counts of COVID-19 could decline when the weather becomes warmer~\cite{xie2020association}. We think the relative growth rate increases in the warmer states because the higher temperature could change how people behave and interact outdoors - people in the warmer states might stay outdoors for a longer period of time increasing the chance of interacting with others. In addition, people might also find it less comfortable to wear a mask in warmer weather which potentially reduces the value in curtailing community transmission and the burden of the pandemic~\cite{eikenberry2020mask}.

\section{Conclusions and Future Work}
Through our analysis of the mined transportation data and the daily confirmed cases COVID-19 infection data, we gain a better understanding of the spread of COVID-19 in the US states along the political lines. We observe that during the quarantine period, when stay-at-home orders were imposed nationally, both the Red and Blue states exhibit a similar level of correlation between their travel patterns and infection growth rates. However post-reopening, we observe that the Red states continue to correlate at a significantly higher level than the Blue states.\\

Upon investigating post-reopening travel and infection data further, we find that both sets of states display similar travel patterns (there are Blue states with increasing travel trends as well). This leads us to conclude that the spread of COVID-19 post-reopening might be less dependent on travel, but perhaps more on the way residents mobilize. This could be related more to safety measures taken while mobilizing, such as mask-usage and social distancing, etc. Another factor explored was that of warmer temperatures in the southern part of the country allowing residents in the Red states to spend more time outdoors interacting with others.\\

There is scope for future work to analyze how these two sets of states acted in previous pandemics and the relationship between the state governments and the federal government. In particular, it would be valuable to quantitatively measure the adherence to the safety measures when people travel in different states. This could be studied in terms of mask-usage, indoor versus outdoor interactions and safety regulations imposed by different states. It would also be valuable to investigate the spread of COVID-19 at the county level, to understand community transmission and socio-economic factors that could help explain the increase in growth rate for Red states.

\bibliography{name}

\end{document}